\documentstyle[manuscript,aps]{revtex}
\input epsf

\newcommand{\beq}{\begin{equation}}
\newcommand{\eeq}[1]{\label{#1} \end{equation}}
\newcommand{\beqar}{\begin{eqnarray}}
\newcommand{\eeqar}[1]{\label{#1} \end{eqnarray}}

\begin{document}
\draft
\preprint{SNUTP-98-15, SOGANG-HEP 233/98, hep-th/9803069}
\setcounter{page}{0}
\title{\Large\bf $SO/Sp$ Monopoles and Branes with Orientifold 3 Plane} 
\author{\large \rm 
Changhyun Ahn\footnote{chahn@spin.snu.ac.kr, Address after June 1,
  1998: Center for Theoretical Physics, Seoul National University,
Seoul 151-742, Korea
}}
\address{ \it Department of Physics, 
Seoul National University, 
Seoul 151-742, Korea} 
\author{\large  Bum-Hoon Lee\footnote{bhl@ccs.sogang.ac.kr}}
\address{ \it Department of Physics, 
Sogang University, 
Seoul 121-742, Korea}
\date{\today}

\maketitle

\begin{abstract}

We study BPS monopoles in 4 dimensional $N=4$ $SO(N)$ and $Sp(N)$ 
super Yang-Mills theories 
realized  as the low energy effective 
theory of $N$ ( physical and its mirror ) 
parallel D3 branes and an {\it Orientifold 3 plane } with
D1 branes stretched between them in type IIB string theory.
Monopoles on D3 branes give the natural understanding by
embedding in $SU(N)$ through the
constraints on both the asymptotic Higgs field ( corresponding 
to the horizontal positions of D3 branes )  
and the magnetic charges ( corresponding to the number of D1 branes) 
imposed by the O3 plane.  
The compatibility conditions of Nahm data for monopoles
for these groups can be interpreted very naturally 
through the D1 branes in the presence of O3 plane.

\end{abstract}

\setcounter{footnote}{0}

\section{Introduction}

In the last year we have seen how string/M theory can be exploited to
understand nonperturbative dynamics of low energy supersymmetric gauge
theories in various dimensions and different number of supersymmetries.
D brane dynamics in string theory is very useful tool for
the gauge 
theory which is realized on the worldvolume of D branes ( See \cite{gk}
for a review and references on the interrelation between D brane dynamics and
the gauge theory). 

Nahm's construction \cite{nahm} of the moduli space of magnetic 
monopoles \cite{manton} was found
in \cite{dia} by considering parallel D3 branes with D1 branes
stretched between them in type IIB string theory. By T-duality along 
the transverse 2 space directions to both D1 and D3 branes this 
configuration leads to D5 and D3 branes which by S-duality will become
NS5 and D3 brane configuration. 
The mirror symmetry of $N=4$ gauge theory in 3
dimensions \cite{hw}
was due to the nonperturbative S-duality of type IIB string theory ( See,
for example, \cite{bo1} ). By T-duality along the 1 space direction, we have
2 parallel NS5 branes with D4 branes suspended between them.
As one changes the relative orientation \cite{bar}
of the two NS5 branes while keeping
their common four spacetime dimensions intact, the $N=2$ supersymmetry
is broken to $N=1$ supersymmetry \cite{egk}. 
By studying this brane configuration
they \cite{egk}
described and checked a stringy derivation of Seiberg's duality. 
This configuration was generalized to the cases with orientifolds which
were used to study $N=1$ supersymmetric gauge theories with gauge
group $SO(N)$ and 
$Sp(N)$ \cite{eva} ( See also \cite{ov} for a relevant geometrical 
approach ). 
%

On the other hand, in the field theory, 
magnetic monopoles have been the object of intense interest as 
solitons which after quantization are complementary
to particles that arise as quanta of the elementary fields.
This has been developed  by Montonen and Olive \cite{mo} that
in certain theories there exists an exact electromagnetic duality
which exchanges solitons and elementary quanta and weak and strong 
coupling. 
In fact,  $N=4$ supersymmetric Yang-Mills theory is the candidate theory
with this duality \cite{osb,bhl}.
In brane picture, the Montonen-Olive duality can be regarded as
S-duality of type IIB string theory in the limit of vanishing of fundamental
string scale. 
In general, it is very hard to solve the general BPS monopole field 
configuration but the systematic construction for the solution has 
been done by 
Nahm \cite{nahm}, so-called
Nahm's equation which is some nonlinear ordinary differential equation.
Up to a coulple of years ago, the origin of Nahm's equation was unclear
mathematical artifact. However, the string theory tells us that
it naturally comes through the D1 brane 
configuration suspended between D3 branes.
This was shown in \cite{dia} for $SU(N)$ monopoles.

In this paper, we generalize the approach of \cite{dia} to $SO(N)$ and
$Sp(N)$ magnetic
monopoles. The $SO/Sp$ monopoles can be understood by embedding the
gauge group into $SU(N)$. The corresponding Nahm's data are obtained
by  imposing some extra constraints to those for $SU(N)$ case.
All these can be naturally realized by putting O3
plane in the D1-D3 brane cofiguration.
The O3 plane allows us to construct the $SO/Sp$ gauge
theories. Furthermore, it provides the natural geometrical origin of
those extra constraints.
in $SO(N)$ or $Sp(N)$ gauge theories. 

\section{Monopoles in supersymmetric gauge theories}
In this section we briefly review the Bogomol'nyi-Prasad-Sommerfield
(BPS)  monopoles \cite{BPS}. 
We consider the model with the Higgs fields $\Phi$ in the adjoint
representation of the arbitrary gauge group $G$ with rank $n $ in the
BPS limit in which the potential of
the scalar fields are ignored. This is the case of the extended
supersymmetric models.


The BPS monopole configurations  can be described by the following
first order equations  
\begin{equation}
B_i = D_i \Phi  \label{BPSeq} 
\end{equation}
where 
$\epsilon_{ijk} B_k = \partial_i A_j -\partial_j A_i +i[A_i,A_j]$
and $D_i \Phi = \partial_i \Phi + [A_i, \Phi]$.   
The asymptotic value $\Phi_0$ of the Higgs field $\Phi$ along some fixed
direction can be written as \cite{weinberg} 
\begin{equation}
\Phi_0 = {\bf h} \cdot {\bf H} 
\label{asymHiggsvalue}
\end{equation}
where the $n$ commuting matrices $H_i$ span the Cartan subalgebra and are
normalized by Tr$H_i H_j=\delta_{ij}$. 
The simple roots can be chosen to
have non-negative inner products with 
the vector $\bf h$ in the root space.  

For the generic value of $\bf h$ with no simple roots
orthogonal to that, the gauge group $G$ is maximally broken to
$U(1)^n$. 
If some simple roots ${\mbox{\boldmath $\gamma$}}_j$ $(j=1,\cdots,k)$
are  orthogonal to $\bf h$,  
then the unbroken gauge group becomes $U(1)^{n-k} \times K$ where $K$ is
the semisimple group corresponding to the root sublattice of
${\mbox{\boldmath $\gamma$}}_j$.
Much physics of this nonabelian unbroken gauge group can be understood
by taking the  
limit of  ${\bf h} \cdot {\mbox{\boldmath $\gamma$}}_j \rightarrow 0 $
of the generic maximal symmetry breaking \cite{lwy}. 
Hence we will mainly consider 
maximal symmetry breaking 
to $U(1)^n$ for simplicity. 

The asymptotic magnetic field can be written as
\begin{equation}
B_i = {\hat r_i \over 4\pi r^2} {\bf g} \cdot {\bf  H}.  \label{Basymp} 
\end{equation}
The quantized magnetic charge ${\bf g}$ is given by
\begin{equation}
{\bf g} = {4\pi\over e} \sum_{a=1}^n m_a 
{\mbox{\boldmath $\beta$}}_a^* 
\label{magcharge} 
\end{equation}
where $ {\mbox{\boldmath $\alpha$}}^* 
 = {\mbox{\boldmath $\alpha$}} / {\mbox{\boldmath $\alpha$}}^2 $ is 
the dual of the root ${\mbox{\boldmath $\alpha$}}$ and the $m_a$
are non-negative integer valued topological charges. 
 

Since each root defines an $SU(2)$
subgroup, we can construct the corresponding $SU(2)$ monopole
solution. 
For maximal symmetry breaking, the monopole corresponding to each of
the $n$ simple roots  
${\mbox{\boldmath $\beta$}}_a$ is the fundamental monopole carrying the
unit of $U(1)$ magnetic charge  and with the mass 
$M_a = {4\pi\over e} {\bf h} \cdot {\mbox{\boldmath $\beta$}_a^*}$.
If some simple roots ${\mbox{\boldmath $\gamma$}}_a$ become orthogonal
to  $\bf h$ giving rise to the unbroken nonabelian gauge group, then
the corresponding monopoles become massless.  
The general
BPS monopoles in Eqs.(\ref{asymHiggsvalue}) and (\ref{magcharge}) 
and the mass $M$ given by
\begin{equation}
M = \sum_{a=1}^n m_a M_a . \label{monopolemass} 
\end{equation}
can be understood as the multimonopole
configuration containing set of  $ m_a$  monopoles of each
of the above $n$ different fundamental monopole types. This
is consistent with the analysis that each fundamental monopole
is associated with four moduli corresponding to the three positions
and one $U(1)$ rotation and that the total dimension of the moduli is
preserved in the limit of some monopoles becoming massless. 

For $SU(N)$, the asymptotic Higgs field can be written as 
\begin{equation}
\Phi_0 = {\bf h} \cdot {\bf H} = 
diag(\mu_1, \mu_2, \cdots,\mu_{N-1}, \mu_N) \label{sunphi}
\end{equation}
with
$\sum_{a=1}^N \mu_a =0$. 
Maximal symmetry breaking corresponds to all different values of
$\mu_a$. 
If some of the $\mu_a$ have the same values, we have some nonabelian
unbroken gauge group. For the maximal symmetry breaking, 
we choose such that $\mu_1 <\mu_2< \cdots
<\mu_{N-1}<\mu_N$. The mass $M_a$ becomes 
\begin{equation}
M_a = {4\pi\over e} {\bf h} \cdot {\mbox{\boldmath $\beta$}_a^*} 
= |\mu_a -\mu_{a+1}| .  \label{sunmass} 
\end{equation}
The magnetic charge of the monopoles, 
${\bf g}\cdot {\bf H}$, becomes 
\begin{equation}
{\bf g}\cdot {\bf H} = diag (k_1, k_2, \cdots, k_{N-1}, k_N)
\label{suncharge} 
\end{equation}
with $k_a = m_a -m_{a-1}$ $(a=1,\cdots,N-1)$, and 
$k_N = -(k_1 +\cdots+k_{N-1})$.

The matrix representation of the Higgs field and the magnetic charges
for other classical gauge groups of $SO(N)$ and $Sp(N)$
can also be done in a straightforward way. 
Related to the later analysis, it is  helpful
to realize these matrices by embedding to $SU(N)$.  
This can be described  with 
the constraints among the $SU(N)$ generators. 
The generators of $Sp(N)$ can be obtained through the constraint 
$ T^t J +JT =0 $ such that $JJ^* = -I$, and  the generators of $SO(N)$ are
given by the constraints $T^t K +KT=0$ with $KK^*=I$. 
The asymptotic values of the Higgs fields in $SO/Sp$ group can then be 
identified to those in $SU(N)$ group satisfying some relations.
Let $\rho_a$ $(a=1, \cdots, n)$ 
represent the magnetic charges of the $SO/Sp$ - multimonopole configuration.
The $SO/Sp$ - magnetic charges will also be related to  
the $SU(N)$ - magnetic charges $m_a$ satisfying some relations.
The results are summarized as follows ( See, for example,\cite{hurt} ).

\vspace{4mm}
\begin{center}
\begin{small}
\begin{tabular}{|c c |c c|}\hline
$G$ & $G$-charges & $\Phi_0$ in $SU(N)$ &
$SU(N)$-charges \\ \hline 
$Sp(N)$ & $\rho_1, \cdots, \rho_n$ & 
$\mu_a=-\mu_{2n+1-a}$ & $m_a=m_{2n-a}=\rho_a$ \\ 
$N=2n$&& $a=1, \cdots, n$ & $a=1, \cdots, n$ \\ \hline
$SO(N)$ & $\rho_1, \cdots, \rho_{n-2}$ &
$\mu_a=-\mu_{2n+1-a}$ & $m_a=m_{2n-a}=\rho_a$ \\
$N=2n$&$\rho_+, \rho_-$ &  $a=1, \cdots, n$ & $a=1, \cdots, n-2$\\
&&&$m_{n-1}=m_{n+1}=\rho_+ +\rho_-$ \\
&&&$m_n=2\rho_+$ \\ \hline
$SO(N)$& $\rho_1, \cdots, \rho_n$ & $\mu_a=-\mu_{2n+2-a}$
& $m_a=m_{2n+1-a}=\rho_a$\\
$N=2n+1$&&$a=1, \cdots, n+1$ & $a=1, \cdots, n-1$\\
&&& $m_n=m_{n+1}=2\rho_n$ \\ \hline 
\end{tabular}
\end{small} \\
\end{center}
\begin{center}
{\bf Table 1}: The embedding of $Sp(N)$ and $SO(N)$ in $SU(N)$. 
\end{center}

\vspace{8mm}

The systematic construction of the general BPS monopole field
configurations can be 
described based on Nahm data \cite{nahm}.
We first review the case of $SU(N)$ monopole. 

The Nahm data for the multi-monopoles carrying charges
$(m_1,\cdots,m_{N-1})$ are defined by the $N-1$ triples of the
analytic $u(m_a)$ valued functions $X^i_a$, where
$i=1,2,3$ and $a=1,\cdots,N-1$, defined on the interval $(\mu_a,
\mu_{a+1})$ and satisfying the following Nahm equations :
\begin{equation}
{d X^i_a \over d s} +{1\over 2} \sum_{j,k=1}^{3} \epsilon_{ijk} 
[X^j_a, X^k_a] = 0. \label{Nahmeq} 
\end{equation}
The Nahm data in two adjacent intervals also
satisfy the following boundary conditions near each $\mu_a$:

i) If $m_{a-1} <m_{a}$, then $X^i_{a-1}$ has a non-zero limit 
$ C^i{}_a = \lim_{s \rightarrow \mu_a^-} X^i_{a-1}$ and 
is analytic at $ s = \mu_a$. Also
$X^i_a$ can be written in a block form expansion
as $ t\equiv s-\mu_a \rightarrow 0$,
\begin{equation}
X^i_a = \left(\begin{array}{cc}
X^i_{a,11} & X^i_{a,12}   \nonumber \\
X^i_{a,21} & X^i_{a,22} 
\end{array} \right) =
\left(\begin{array}{cc}
C^i_a +{{\cal O}(t)} &   {{\cal O}(t^\gamma)} \nonumber \\
{{\cal O}(t^\gamma)} &  {T^i_a \over {s-\mu_a}}      
\end{array} \right) . \label{bdrycond}
\end{equation}

where $T^i_a$ forms an
$(m_a-m_{a-1})$ - dimensional irreducible representation of $su(2)$
and $\gamma = (m_a-m_{a-1}-1)/2$. The upper diagonal block is analytic
and the lower diagonal block is meromorphic in $t$.   

ii) If $m_{a-1} > m_{a}$, the roles of $m_{a-1}$ and $m_a$ are reversed. 

iii) If $m_{a-1} = m_{a}$,  $X^i{}_{a-1}$ and  $X^i{}_a$ are both analytic
near $s=\mu_a$ with finite limits required to satisfy a certain
regularity condition.

We take the convention of $m_0=0=m_N$ to cover the boundary conditions at
$\mu_1$ and $\mu_N$.

As for the $SO(N)$ and $Sp(N)$ monopoles, the Nahm data are naturally
described in relation to the embedding of these gauge groups into
$SU(N)$. 
The  $SO(N)$ and $Sp(N)$ multimonopole magnetic charges $\rho_a$ and the
asymptotic Higgs fields will correspond to $SU(N)$ charges $m_a$ and Higgs
fields as given in the Table 1. With this correspondence, the Nahm
data for $SO(N)$ and $Sp(N)$ monopoles will satisfy the same Nahm
equation in Eq.(\ref{Nahmeq}) as that of $SU(N)$. As for the boundary
conditions between two adjacent intervals, we need the following one
more set of conditions in addition to those for the $SU(N)$ described above
\cite{nahm,hurt}. 
There exist matrices $C_a$ satisfying
\begin{equation}
T^i_{N-a}(-s)^t = C_a T^i_a(s) C_a^{-1}
\label{cmatrix}
\end{equation} 
with
$C_{N-a}=C_a^t$ for $Sp(N)$ and $C_{N-a} = -C_a^t$ for $SO(N)$.

The physical origin of the equivalent description of the monopoles
based on the Nahm's equation with special boundary conditions are not
clear at the field theory level. We will see
in the next section how this relation is naturally realized through
the D branes.

\section{$SO(N)$ and $Sp(N)$ Monopoles and Branes with Orientifold }

The brane dynamics at low energy of parallel $N$ D3 branes stretched in
$(x^0, x^1, x^2, x^3)$ directions 
in type IIB string theory is described by four dimensional $N=4$ 
$SU(N)$ supersymmetric gauge theory. 
Let us consider, as in Figure 1, $N$ parallel D3 branes
with $x^6$ values $\mu_1 < \mu_2 <  \cdots < \mu_{N-1} <  \mu_{N}$ 
and  $m_a$ ($a=1,\cdots,N-1$) D1 branes connecting $a$-th and 
the $a+1$-th D3 branes. 
On the D3 brane worldvolume,
this configuration describes the multimonopole configuration in 
the maximally broken $SU(N)$ gauge theory.
The Abelian $U(1)$ factor corresponds to the center of mass dynamics of
the D3 branes, which is decoupled. Thus we consider $SU(N)$ gauge
theory instead of $U(N)$.
This is realized by the extra condition $ \sum_{a=1}^{N} \mu_a =0$.
The $x^6$ coordinates $\mu_a$ of D3 branes  
are identified with the asymptotic Higgs field value $\Phi_0$ in
Eq.(\ref{sunphi}) in Section II. The D1 branes and their low energy
dynamics  will be
described as the $SU(N)$ multimonopoles with the magnetic charge 
$( m_1, \cdots, m_a, \cdots, m_{N-1} ) $ and their moduli .
The net magnetic charge, $k_a$, 
induced on the $a$-th D3 brane coming from
the $m_{a-1}$ D1 branes in the interval $(\mu_{a-1}, \mu_{a})$ and
the $m_{a}$ D1 branes in the interval $(\mu_{a}, \mu_{a+1})$ will be
given by the difference $k_a = m_a-m_{a-1}$. 
This magnetic charges of the multimonopole configuration in branes
exactly corresponds to those in Eq. (\ref{suncharge}) in
field theory description.

\begin{center}
\leavevmode
\epsfxsize=5.0in
\epsfbox{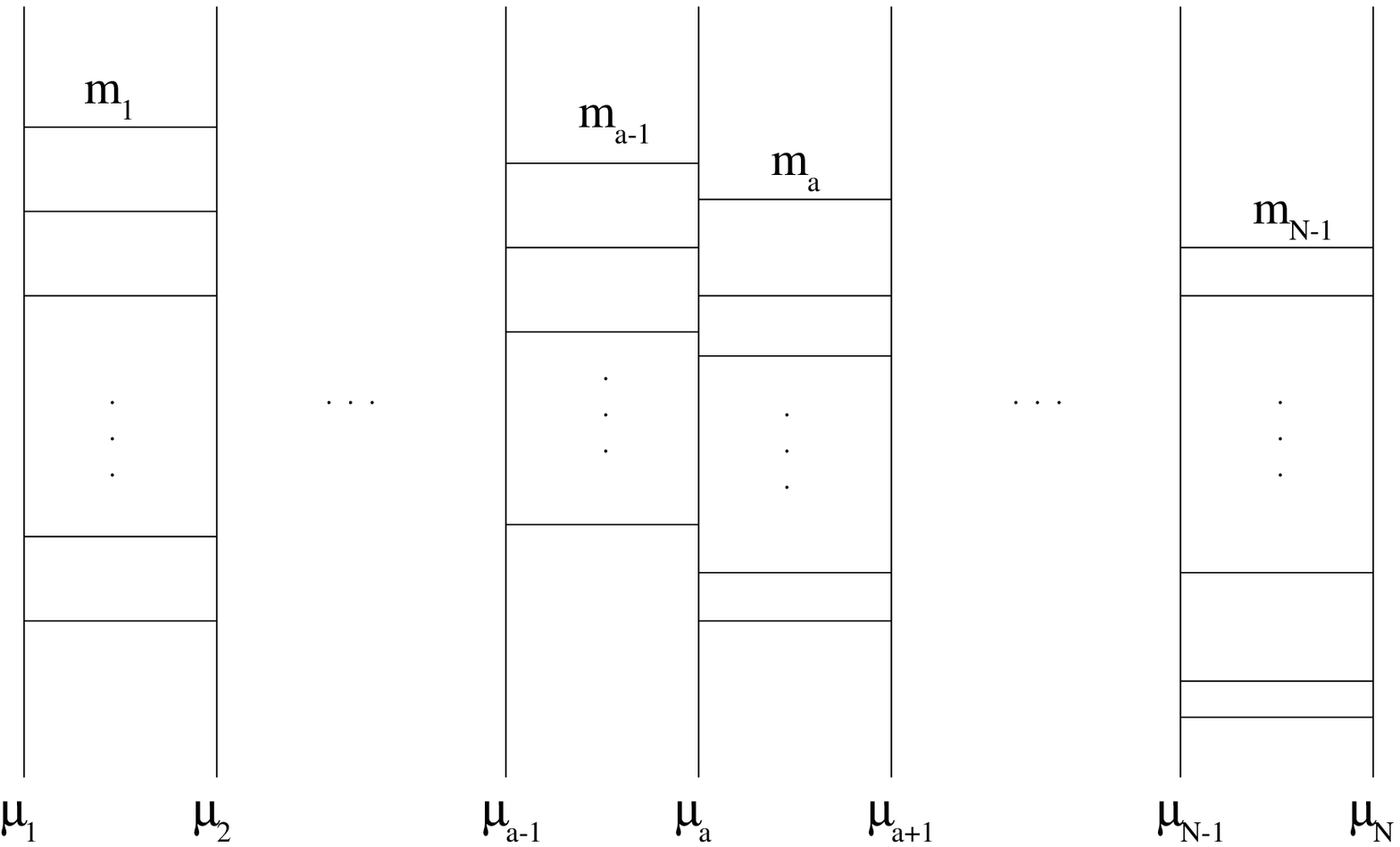}
\end{center}

${\bf Figure 1}$: Brane configuration for $SU(N)$ monopoles.  D3
branes are represented by the vertical lines with space coordinates  
$(x^1, x^2, x^3)$, and D1 branes by the horizontal lines in $x^6$
direction.  The $m_a$'s are the number of D1 branes and $\mu_a$'s  the
$x^6$ positions of D3 branes. Generically, the values of $\mu_a$'s and
$m_a$'s and the positions of D1 branes along the vertical directions
are arbitrary.

\vspace{0.5cm}

From the point of view of D1 branes, the above moduli can be
understood as the moduli describing the vacua of the 1+1 dimensional
field theories of D1 branes stretched between the infinitely heavy D3 branes.
If the $m$ D1 branes were with infinite lengths without D3 branes, the
low energy 
effective action would be the 1+1  
dimensional supersymmetric $U(m)$ nonabelian gauge theory with worldvolume in
$(x^0,x^6)$ which can be obtained by  
the dimensional reduction  from 10 dimensional $N=1$ super
Yang-Mills action.
The theory on D1 branes has eight supercharges, and there are 2 dimensional
gauge fields, scalars and their superpartners. The scalars come from the transverse
oscillations to the worldvolume of D1 branes. There are two types of
scalars : $X^i$ $(i=1,2,3)$ along the D3 brane  and the rest 
$X^\mu$ $(\mu=4,5,7,8,9)$ which are transverse to both D1 and D3
branes. 
Now consider the D1 branes ending on the
D3 branes at $\mu_a$ and $\mu_{a+1}$.  We will use the subscript $a$
for the corresponding quantities, e.g., $X^i_a$ and $X^\mu_a$.
We can choose the gauge field to be zero by a gauge choice (except the
Wilson line phase). In other worlds, the gauge groups are frozen and
become essentially the global symmetry.
The condition for supersymmetric ground states for D1 brane
configurations can be easily obtained from the effective action.
The ending of D1 branes on the D3 branes at $\mu_a$ and
$\mu_{a+1}$ require some boundary conditions to the fields
\cite{dia}. 
The boundary conditions of $X^{\mu}_a, \mu=4, 5, 7, 8, 9$ are as follows:
$X^{\mu}_{a-1}$ and $X^{\mu}_{a, 11}$ are analytic in a neighborhood
of $s=\mu_a$ with finite limits as $s \rightarrow \mu_a$. This condition
is applied for $X^{i}_a$ as well. 
The parameter $s$ is used to represent the value of $x^6$ along the D1
branes for comparison with the notations in Section II.
The fields
$X^{\mu}_{a, 12}, X^{\mu}_{a, 21}$ and $X^{\mu}_{a, 22}$ are bump fields
which are compactly supported away from $s=\mu_a$.  
With these conditions imposed, the equations for the $ X^{i}_{a}$
fields reproduces the Nahm's equation (\ref{Nahmeq}).

To see the detail of the boundary condition for 
the fields $X^i_a$, let us assume that $m_{a-1}<m_a$. Then we write
$X^i_a$ in the block $2\times 2$ block form as
Eq.(\ref{bdrycond}). Then the boundary conditions can be seen to be
exactly the same as those described in Section II. That is,
$X^i_{a, 11}$ is analytic and has the well defined limit, and $X^{i}_{a,  12}$
and $X^{i}_{a, 21}$ are analytic 
around $s=\mu_{a}$ up to some order 
$\gamma$. Moreover, $X^{i}_{a, 22}$ 
is in the meromorphic form $T^i_a/(s-\mu_a)$ where $T^i_a$ is the
irreducible $SU(2)$ representation. Thus D1 brane point of view shows how
the Nahm's equation and its boundary conditions 
naturally occur.
Based on the correspondence between brane configuration and field
theory configuration for the $SU(N)$ monopole we now move on to
$Sp(N)$ and $SO(N)$ gauge groups. 
   
Understanding the $Sp(N)$ and $SO(N)$ monopoles by embedding these groups
in $SU(N)$ will
correspond to inserting the orientifold 3 plane to the previous brane
configuration of 
$N$ parallel D3 branes with D1 branes connecting them.
To keep the supersymmetry the orientifold 3 plane should be inserted
parallel to D3 branes . 
Also when we put orientifold 3 plane at some position,
the same amount as  ``physical'' D branes appearing in
the left hand side of it should be present in the right hand side of
it at the opposite positions as their images. 
This requires to put the orientifold 3 plane
at the central position of $N$ D3 branes.
  
For an $O3_{+}$ plane of
positive charge with $1/2$ D3 brane charge, the gauge group is
$Sp(N)$ while for a $O3_{-}$ plane of negative charge carrying $-1/2$
D3 brane charge, the gauge group becomes $SO(N)$ ( See, for example,
\cite{egkt} ). We will consider how
O3 plane plays an important role for getting the constraints on the
$\mu_a$ and the magnetic charge $m_a$ and see  
the compatibility conditions of Nahm data for monopoles
for these groups very naturally 
through the D1 branes in the presence of O3 plane.

\subsection{$Sp(N)$: $N = 2n $}

Now we put $O3_{+}$ 
in parallel to D3 branes into the central position 
between $N/2$-th D3 brane $N/2+1$-th D3 brane,  as in Figure 2. 
The $x^6$ position of the orientifold will then be $x^6=0$. 
The presence of the orientifold requires that 
the brane configuration to the right hand side should be restricted to
be the mirror image of those in the  left hand side.
As for the positions of D3 branes, this means that 
\begin{equation}
 \mu_1=-\mu_{N}, \cdots, \mu_{N/2}=-\mu_{N/2+1}. \label{spphi}
\end{equation}
As for the number of D1 branes which corresponds to the  magnetic
charges, the orientifold 
projection requires  
\begin{equation}
 m_1=m_{N-1}, \cdots, m_{N/2-1}=m_{N/2+1}, m_{N/2} \label{spmag}.
\end{equation}
There are only $n$ independent $m_a$ ($a=1,\cdots,n$) and these are
identified  with the $Sp(N)$ magnetic charges $\rho_a$.
Eqs.(\ref{spphi},\ref{spmag}) are exactly coincident 
with the constraints of embedding $Sp(N)$ in $SU(N)$ 
we have described in Section II and given in Table 1.
Note that the group theoretical constraints of embedding now has
very natural geometrical interpretation in terms of both the
positions of D3 branes and the number of D1 branes. 

For the observer on D3 branes,  it describes a moduli space of $Sp(N)$
monopoles while
for the observer on D1 branes, it leads to a moduli space of vacua
of the field theory on D1 branes with $SO$ symmetry. As mentioned
before the gauge group will be frozen and becomes a global symmetry.
For those stretched between D3 branes in the interval 
$(\mu_a, \mu_{a+1})$, the group will be $SO(m_a)$. 
Without the orientifold, the group would be $U(m_a)$ leading to
the Nahm equation and the boundary conditions the same as those
studied in $SU(N)$ monopoles in the above. 
The presence of the orientifold will lead to further projections. 

To be more specific, there are $m_{a-1}$ D1 branes to the left of the
D3 brane at $\mu_a$, and $m_a$ D1 branes to the right. Due to the
orientifold, there exist $m_{a-1}$ to the right and $m_a$ to the left
at $-\mu_a$ as in Eq.(\ref{spphi},\ref{spmag}). 
Let us assume that $m_{a-1} < m_a $. Then, there will be net 
$k_a = m_a -m_{a-1}$ D1 branes ending from the right at $\mu_a$
and the same numbers from the left at $-\mu_a$. Without the
orientifold, all the states with the $U(k_a)$ charges at end points of the D1
branes  at both boundaries  will be allowed. 
With the orientifold, they are no longer
independent. Only those states invariant under the world sheet parity 
will be survived. 
Under the worldsheet parity $\Omega$, the Chan-Paton states $|ij>$ will be
transformed into $(C_a)_{ii'}|i'j'> (C^{-1}_{N-a})_{j'j}$. 
We omitted transformation of other irrelevant state indices.
Equivalently the world sheet parity action can be rewritten in terms of 
the action of $C_a$ and $C_{N-a}$ on $T_a$ and
$T_{N-a}$ as given in Eq.(\ref{cmatrix}).  
Acting $\Omega$ twice will leave the constraints on $C_a$ and $C_{N-a}$ as 
symmetric projection $C_{N-a} = C^t_a$ for $SO(k_a)$  group, and
antisymmetric projection $C_{N-a} = - C^t_a$ for $Sp(k_a)$  
group \cite{gp}. Note that  $SO(k_a)$ ($Sp(k_a)$) group for D1 branes
corresponds to  $Sp(N)$ ($SO(N)$) group on D3 branes. 
To summarize, we have shown that the extra symmetry in the Nahm data
for $Sp/SO$ groups coming from the constraint in
Eq.(\ref{cmatrix}) arises in the brane configuration through the
invariance of the states under the world sheet parity.

The simplest example is $Sp(2)$. 
The equivalence of $Sp(2)$ and $SU(2)$ arises because there is no
states projected out by the orientifold.
When one considers $Sp(4)$, there exist two kinds of monopoles 
characterized by $m_1(=m_3)$ and $m_2$ while for $SU(4)$ there are
three independent monopoles by $m_1, m_2, m_3$ since all states with
different $m_1$ and $m_3$ charges in $SU(4)$ configurations are
projected out by the orientifold. For higher $N \geq 6$,
it is straighforward to analyze this procedure as well.

\begin{center}
\leavevmode
\epsfxsize=5.0in
\epsfbox{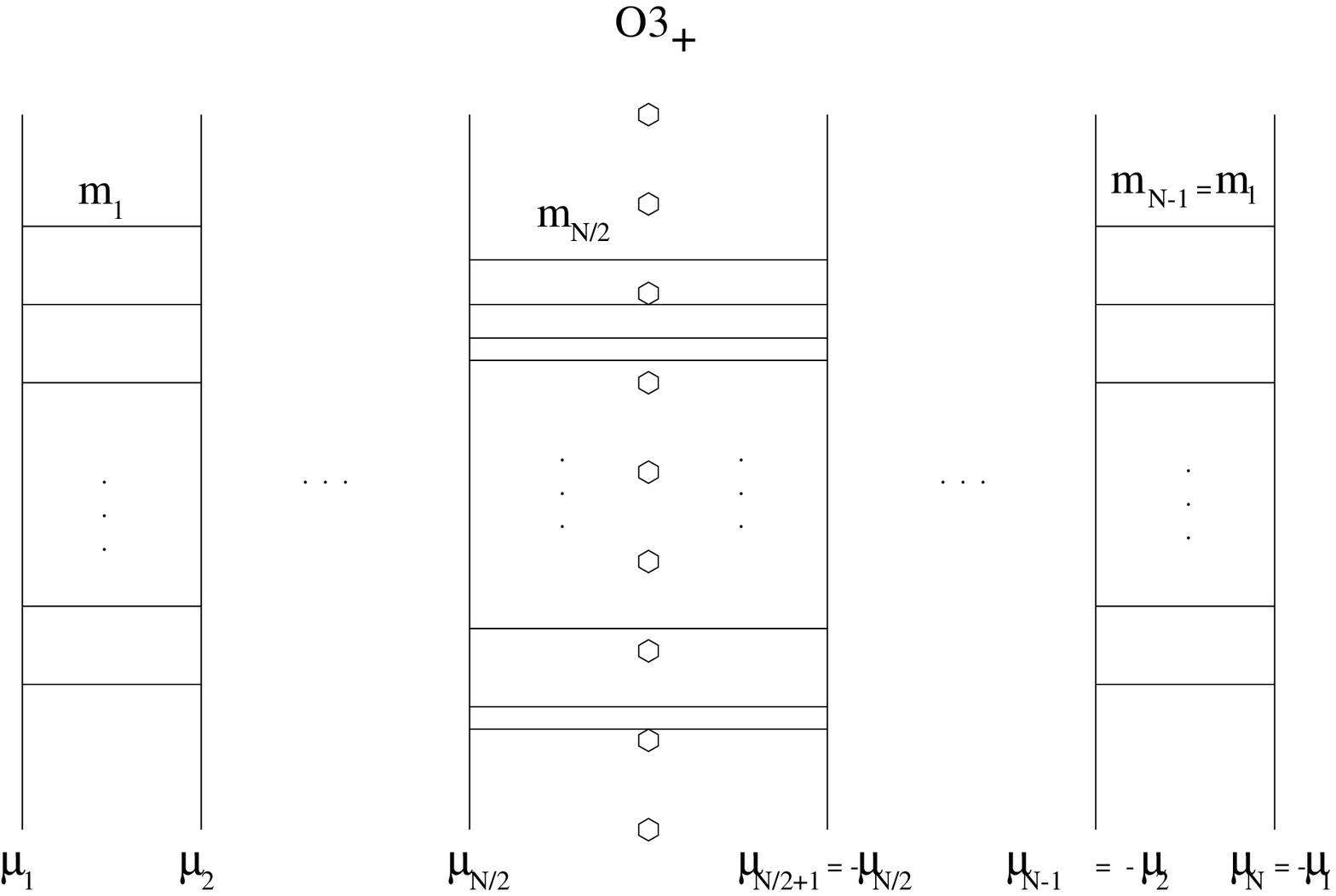}
\end{center}

${\bf Figure 2}$: Brane configuration with orientifold
$O3_{+}$ denoted by the hexagons for  $Sp(N)$ monopoles.
The values of $\mu_a$'s and $m_a$'s appearing in the right hand side
of orientifold are strictly restricted according to the orientifold
projection.
 
\vspace{0.5cm}

\subsection{$SO(N)$: $N=2n+1$}

In this case, we
consider $O3_{-}$, as in Figure 3,
rather than $O3_{+}$  at the position
of $x^6=0$ where a single D3 brane gets stuck on the $O3_{-}$
without its mirror image. 
As in previous case, the orientifold action gives rise to the 
following conditions,
$\mu_1=-\mu_{N}, \cdots, \mu_{(N-1)/2}=-\mu_{(N+3)/2}, \mu_{(N+1)/2}=
-\mu_{(N+1)/2}$ and $m_1=m_{N-1}, \cdots, m_{(N-3)/2}=m_{(N+3)/2},
m_{(N-1)/2}=m_{(N+1)/2}$
which exactly matches the embedding constraints of $SO(N)$ in $SU(N)$
as shown Table 1. We obtained these group theoretical constraints as a
natural geometrical properties of the orientifold.
Montonen-Olive electric-magnetic duality in field theory tranforms
$SO(2n+1)$ electric charges
into $Sp(2n)$ magnetic charge and vice versa.  This can be regarded as
S-duality of type IIB string theory in the limit of string scale, $l_s 
\rightarrow 0 $. It turns out that a system of $O3_{-}$ and D3
brane stuck on it transforms as $O3_{+}$ under the S duality of type
IIB string theory.
As before, the effective theory on D3 branes will describe the $SO(N)$
monopole moduli while the theory on  D1 branes between 
the interval $(\mu_a, \mu_{a+1})$ through the 1+1 dimensional
$Sp(m_a)$ supersymmetric theory give the
monopole description through the Nahm data. 
The analysis of this Nahm data goes in parallel with the $Sp(N)$
case. The only difference is that 
states that
are invariant under the antisymmetric projection $C_{N-a} = -C_a^t $
will survive because
the group  is $Sp(k_a)$ instead of $SO(k_a)$.  
Again we obtained this exact 
correspondence to the symmetry constraints in Eq. (\ref{cmatrix}) via
the geometrical world sheet parity transformation properties.
      
For the simple $SO(3)$ case, the monopole is described by $m_1(=m_2)$
which implies  $SU(2)$ monopole all other $SU(3)$ states being
projected out. This can be also understood as dual to $Sp(2)$ theory
we have discussed in previous subsection under the Montonen-Olive
duality or S-duality in type IIB string theory. 
For higher $N \geq 5$ case, we can understand in a similar way.

\begin{center}
\leavevmode
\epsfxsize=5.0in
\epsfbox{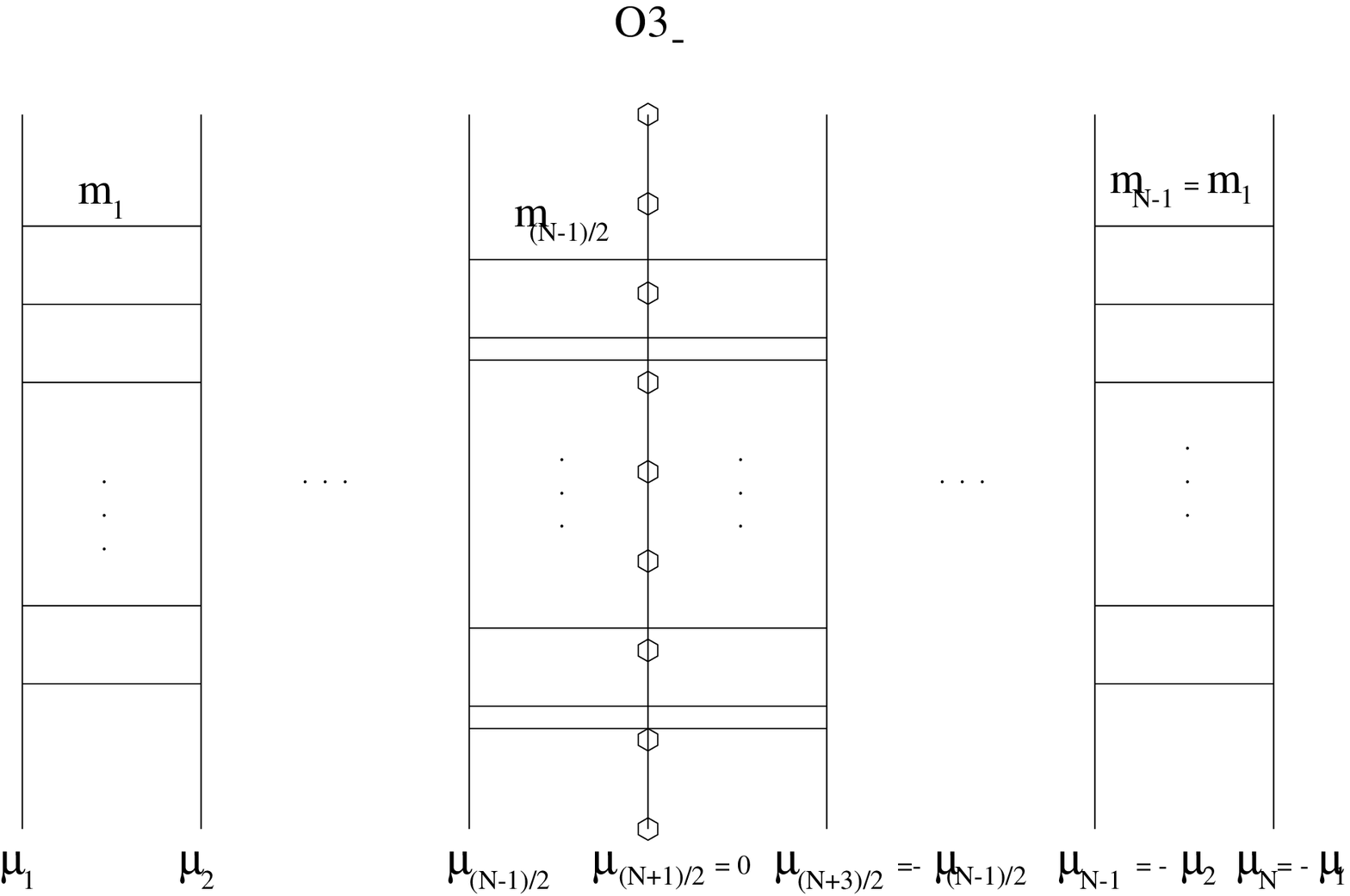}
\end{center}

${\bf Figure 3}$:  Brane configuration with orientifold
$O3_{-}$ denoted by the hexagons for  $SO(N)$ monopoles. A single D3
brane is stuck on the $O3_{-}$ plane.
The values of $\mu_a$'s and $m_a$'s corresponding to the images 
are also restricted according to the orientifold projection.

\vspace{0.5cm}

\subsection{$SO(N)$: $N=2n$}

Now let us put $O3_{-}$, as in Figure 4,
parallel to D3 branes into the central position 
between $N/2$-th D3 brane $N/2+1$-th D3 brane
at $x^6=0$. Once again, the orientifold action leads to
the following conditions, $\mu_1=-\mu_{N}, \cdots, \mu_{N/2}=-\mu_{N/2+1}$
and $ m_1=m_{N-1}, \cdots, m_{N/2-1}=m_{N/2+1}, m_{N/2}$, which is the
same conditions as the embedding of $SO(N)$ in $SU(N)$ shown in Table 1.
This is the same as the case of $Sp(N)$ except that there is a
negative $O3_{-}$ plane rather than positive one.
The $SO(N)$ gauge theory with $N$ even is self dual under the Montonen-Olive
duality. This corresponds to the fact that $O3_{-}$ is self dual under
the S duality. 
We now describe the $SO(N)$ gauge theory on D3 brane from the D1 brane
point of view that leads to the relation with the Nahm data. 
The analysis is the same as previous $SO(2n+1)$ case. 
The geometrical world sheet parity transformation properties
lead us to the antisymmetric projection $C_{N-a} = -C_a^t $, which is
the same as the symmetry constraints of Nahm data in Eq. (\ref{cmatrix}). 
     
Notice that there are no nonsingular monopoles for $SO(2)$ which can
be interpreted as due to the fact that the fundamental strings between D3
branes which obtained by S-duality 
are projected out when we have $O3_{-}$ plane. 
For higher $N \geq 4$ case, it is straightforward to analyze according
to the above projection description.
 
\begin{center}
\leavevmode
\epsfxsize=5.0in
\epsfbox{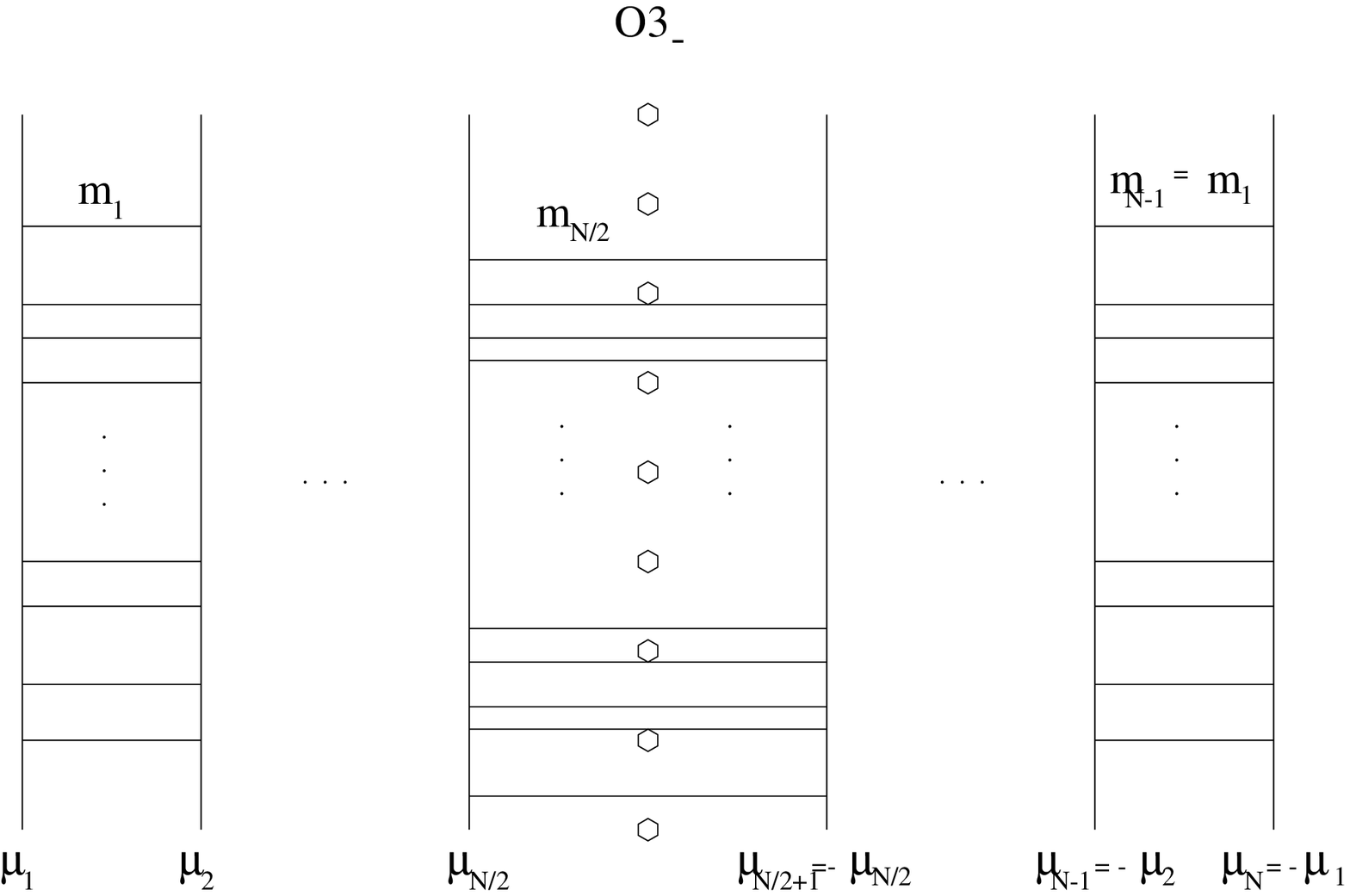}
\end{center}

${\bf Figure 4}$: Brane configuration with orientifold
$O3_{-}$ denoted by the hexagons for $SO(N)$ monopoles.
The images are constrained as above by orientifold projection.

\vspace{0.5cm}

\section{Discussion and Conclusion}

We have studied the correspondence between the brane web of D3 branes and
D1 branes with orientifold in type IIB string theory and the Nahm's data 
for  the general BPS field configuration which can be summarized as

${\bullet}$ The asymptotic Higgs fields,
${\mu_a}$ $\leftrightarrow$ the $x^6$ position of D3 brane.

$\bullet$ The magnetic charge of fundamental monopole, 
$m_a$ $\leftrightarrow$ the number of D1 branes stretched between two D3
branes. 

These are already known for $SU(N)$ monopoles and the new things for 
$SO(N)$ and $Sp(N)$ are as follows.
 
$\bullet$ The constraints on $\mu_a$ and $m_a$ from the embeddings of 
$SO(N)$ and $Sp(N)$ in $SU(N)$ $\leftrightarrow$ O3 plane projection.

$\bullet$ The existence of matrices $C_a$ in Eq.(\ref{cmatrix}) 
$\leftrightarrow$ O3 plane projection ( the worldsheet parity ) 
of D1 branes.

As we have seen, O3 plane plays the crucial role in understanding the $SO/Sp$ 
monopoles in brane configuration providing the natural physical origin of the 
Nahm's equation and its boundary conditions, which was mysterious at the 
field theory level.

The moduli for the multimonopoles also have a natural interpretation in brane 
configuration as the Wilson loop and the motion of the D1 branes. Considering 
the recent progress of the metrics in the moduli space in various aspects 
in the field theories \cite{lwy}, it is interesting to describe these  
and to study the monopole scattering in detail in terms of branes.

\vspace{2cm}

\centerline{\bf Acknowledgments} 

We thank the participants and lecturers of the Second  Winter School, where
this work has been motivated,
on {\it Branes, Fields and Mathematical Physics } 
Feb. 9 - 20, 1998, APCTP, Seoul, Korea for useful discussions.
We also thank E.J. Weinberg for discussions. This work was supported
by the Korean Science and Engineering Foundation through Center for 
Theoretical Physics and by the Korean Ministry of Education (BSRI-97-2414).

\end{document}